\documentclass[%
 reprint,
superscriptaddress,
 amsmath,amssymb,
 aps,
]{revtex4-2}

\usepackage{graphicx}
\usepackage{dcolumn}
\usepackage{bm}
\usepackage{scalerel}
\newcommand\scht[1]{\stretchrel*{$\textsc{#1}$}{\textsc{x}}}
\usepackage{braket}
\usepackage{siunitx}
\usepackage{nicefrac}
\usepackage{graphicx}
\usepackage{dcolumn}
\usepackage{bm}
\usepackage{xcolor}
\usepackage[makeroom]{cancel}
\usepackage{mathtools}

\usepackage{comment}

\usepackage{enumitem}
\usepackage{interval}
\usepackage{amsmath,amssymb}
\usepackage{interval}
\usepackage[flushleft]{threeparttable}
\usepackage{booktabs}
\usepackage{siunitx}
\usepackage{textgreek}
\usepackage{lineno}
\DeclareSIUnit\barn{b}

\begin{document}

\preprint{APS/123-QED}

\title{Microstructured plastic scintillators for beam profiling in medical accelerators}

\author{Veronica Leccese}
\affiliation{
 Institute of Physics (IPhys), Laboratory for Ultrafast Microscopy and Electron Scattering (LUMES), École Polytechnique Fédérale de Lausanne (EPFL), Lausanne 1015 CH, Switzerland.\unpenalty~}
\author{Michele Caldara}
\affiliation{
 Institute of Physics (IPhys), Laboratory for Ultrafast Microscopy and Electron Scattering (LUMES), École Polytechnique Fédérale de Lausanne (EPFL), Lausanne 1015 CH, Switzerland.\unpenalty~}
\author{Samuele Bisi}
\affiliation{
 Institute of Physics (IPhys), Laboratory for Ultrafast Microscopy and Electron Scattering (LUMES), École Polytechnique Fédérale de Lausanne (EPFL), Lausanne 1015 CH, Switzerland.\unpenalty~}
\author{Marcello Pagano}

\affiliation{
 Institute of Physics (IPhys), Laboratory for Ultrafast Microscopy and Electron Scattering (LUMES), École Polytechnique Fédérale de Lausanne (EPFL), Lausanne 1015 CH, Switzerland.\unpenalty~}

\author{Simone Gargiulo}
\affiliation{
 Institute of Physics (IPhys), Laboratory for Ultrafast Microscopy and Electron Scattering (LUMES), École Polytechnique Fédérale de Lausanne (EPFL), Lausanne 1015 CH, Switzerland.\unpenalty~}

\author{Carlotta Trigila}
\affiliation{
Department of Biomedical Engineering, University of California, Davis, Davis, CA, United States.\unpenalty~}

\author{Arnaud Bertsch}
\affiliation{
  Institute of Electrical and Micro Engineering (IEM), Laboratory of Microsystems (LMIS4), École Polytechnique Fédérale de Lausanne (EPFL), Lausanne 1015 CH, Switzerland.\unpenalty~}

\author{Alessandro Mapelli}
\affiliation{
Institute of Mechanical Engineering (IGM), Advanced NEMS Laboratory (NEMS), École Polytechnique Fédérale de Lausanne (EPFL), Lausanne 1015 CH, Switzerland.\unpenalty~\\
}
\affiliation{
Confovis GmbH, Jena, Germany.\unpenalty~\\
}

\author{Fabrizio Carbone}\email{fabrizio.carbone@epfl.ch}
\affiliation{Institute of Physics (IPhys), Laboratory for Ultrafast Microscopy and Electron Scattering (LUMES), École Polytechnique Fédérale de Lausanne (EPFL), Lausanne 1015 CH, Switzerland.\unpenalty~
}

\date{\today}

\begin{abstract}
A novel beam profiler based on microstructured scintillation resin is presented. The detector consists of a bundle of waveguides, with an active area of 30 $\times$ 30 mm$^{\mathrm{2}}$ and a pitch of 400 \textmu m, obtained by molding a scintillating resin into a microfabricated PDMS mold. A first prototype,  coupled to an array of photodiodes and readout electronics, which potentially allows profile rates of more than 7 kHz, has been tested using both a UV source and a proton beam accelerated at different energies, such as those typically used in proton therapy. 
The results obtained during the experimental test campaigns were compared with theoretical simulations showing a good agreement with the modeling expectations, thus confirming the validity of this novel design for microstructured scintillating detectors.
\end{abstract}

\maketitle

\section{\label{sec:intro}Introduction}
Charged particle therapy represents the most advanced form of radiotherapy \cite{LaRiviere2019, Mohan2017}. The peculiarity of the use of hadrons in radiotherapy is the deposited energy per unit track in the body, \textit{i.e.}, Bragg peak, which makes them the perfect tool to localize energy deposition in the patient tumor, minimizing the dose delivered to surrounding healthy tissues and organs \cite{Newhauser2015, Smith2006}.\\
Beam instrumentation and diagnostics play an important role notably in the hadron-therapy facilities, where the complexity of the accelerators and the transfer lines transporting the beam from the accelerator to the treatment rooms necessitates several diagnostic tools along the beam path in addition to dose delivery measurement systems. Moreover, Beam Diagnostics (BD) is essential to commission beams in a new accelerator and it allows logging beam parameters during the machine quality assurance (QA), thereby helping the identification of slow drifts or discrepancies from reference settings.\\
Beam transverse profile is one of the key beam parameters; it serves to set the correct optics in the accelerator and, in conjunction with quadrupoles and dipoles, it is used to measure the beam emittance, energy, and energy spread \cite{STREHL2006}. Until today, beam profile measurements at high ion energies remain a challenge due to the high dose to which materials are exposed. 
During the past decades, many solutions have been proposed and refined, ranging from wire-based technologies (profile grids or wire scanners) \cite{Arutunian2005, Taylor2016}, gas-ionization-based devices (residual gas monitors \cite{Levasseur2017}) to scintillating material-based devices (scintillating screens or fibers \cite{Blumer1995}). In recent years, scintillating fibers or, more generally, plastic scintillators connected to photodetector and readout systems turned out to be a simple and reliable solution for beam diagnostics \cite{Loch2020,kim2022three}. The drawbacks of such well-known detectors are their limited lifetime due to permanent radiation damage on the fibers most exposed to the beam \cite{Wetzel_2022} and the spatial resolution, constrained by the dimension of the fiber itself. Moreover, assembling many fibers with a good alignment is a laborious and complex process, especially if a large area needs to be covered.
Microchannels filled with scintillating liquid have been proposed as an alternative to scintillating fiber assemblies \cite{Mapelli2010sensors,Mapelli2010,Mapelli2011ScintillationPD,Maoddi2014}, however the difficulty to homogeneously fill the microchannels together with the possibility of having liquid leaks, make those devices barely suitable for high vacuum applications. 
Recently, a CMOS detector has been successfully employed in air, monitoring a pencil proton beam scanned source, becoming a promising alternative for monitoring beam transverse profiles \cite{FLYNN2022166703}. For what concerns minimally invasive detectors, especially interesting for circular accelerators, considerable advances have been reported on residual gas \cite{Levasseur2017} or gas curtain \cite{Welsch2022} devices. \\
Here, we present a microfabricated scintillating resin-based detector able to achieve micrometer resolution and overcome radiation damage issues; the device is intended for transverse profile monitoring in particle accelerators and it aims to address the limitations associated with the existing beam profilers.

\section{\label{sec:physical} Microscintillator beam profiler}
\begin{figure}[htbp]
\begin{center}
\includegraphics[width=\linewidth]{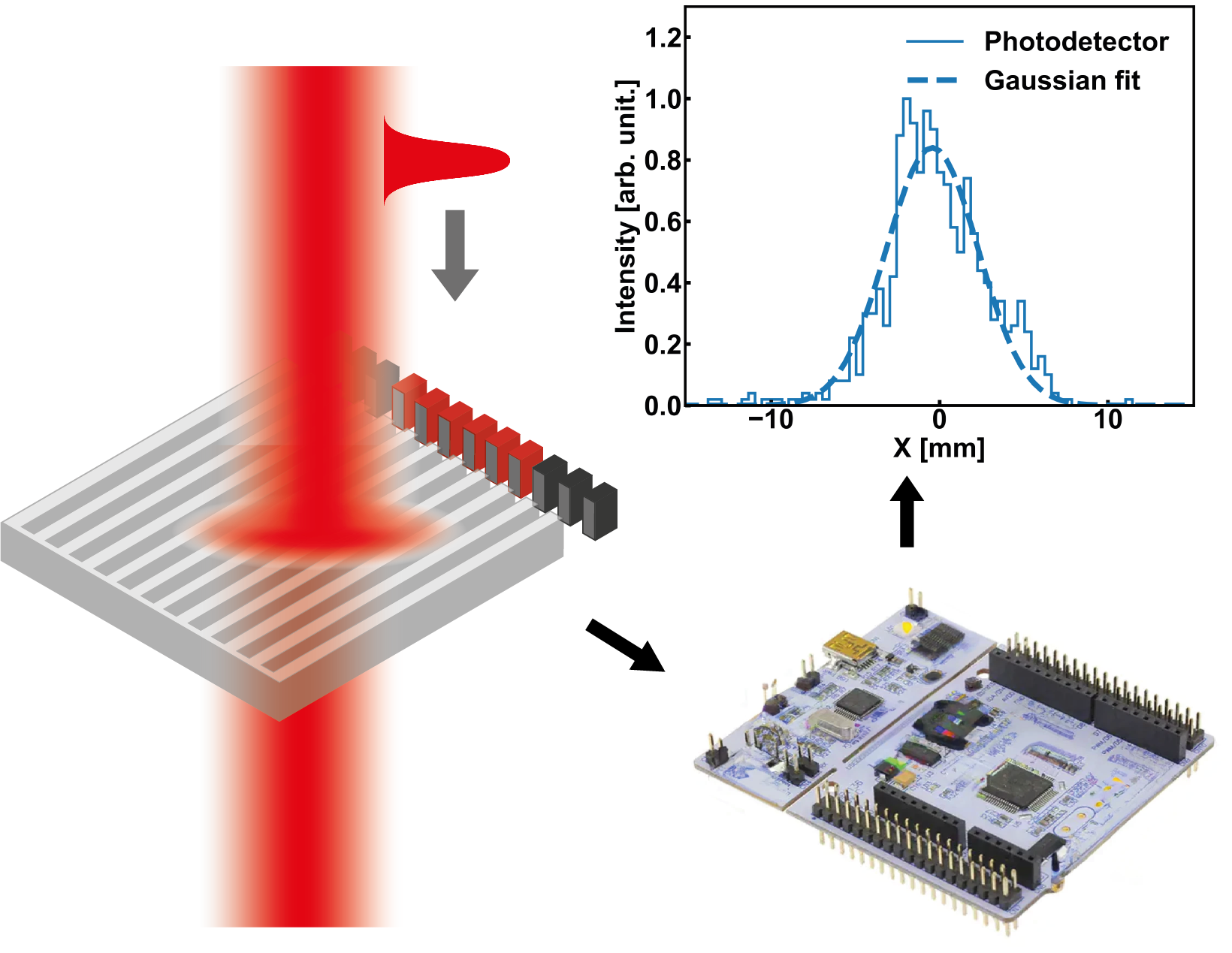}
\caption{Sketch of the beam profiler working principle. The proton beam hits the scintillating active area of the beam profiler. The photons generated by the scintillation process are collected by the PD array connected at each channel end. The signal is acquired and converted into photocurrent, by which the beam profile can be reconstructed knowing the PD array geometry.}
\label{sketch}
\end{center}
\end{figure}

The working principle of the microscintillator beam profiler consists of an active area completely made of scintillating resin, which emits photons when struck by a particle beam.
The active area comprises 75 channels, each with a square cross-section 200 \textmu m $\times$ 200 \textmu m, a length of 30 mm, and separated from one another with a pitch of 400 \textmu m. Thus, the total active area of the full device is 30 $\times$ 30 mm$^{\mathrm{2}}$, which is compatible with the typical beam transverse dimensions in accelerators. The 400 \textmu m-pitch allows reaching a theoretical resolution of $\sim$115 \textmu m, which can be further reduced by decreasing the pitch size \cite{wang2018ultimate,webel1996r}. Thanks to the different refractive indices of the scintillating resin ($R=1.58$) and the surrounding air/vacuum, the channels act as a waveguide for the photons as long as they hit the resin-air interface with an angle smaller than the critical one (39.2°). The photons are guided toward the array of photodetectors (PDs) which has the same pitch as the active area and is placed at a 1 mm distance in front of the end of the channels. The PDs detect and convert the photons into an electrical signal, from which the profile of the beam can be retrieved. The signal obtained from the PD array is retrieved using a microcontroller integrated into a custom-made printed circuit board (PCB). 
Fig.~\ref{sketch} shows the working principle of the device. 

\subsection{\label{sec:modelling} Modelling and Simulations}
\begin{figure}[b!]
\begin{center}
\includegraphics[width=\linewidth]{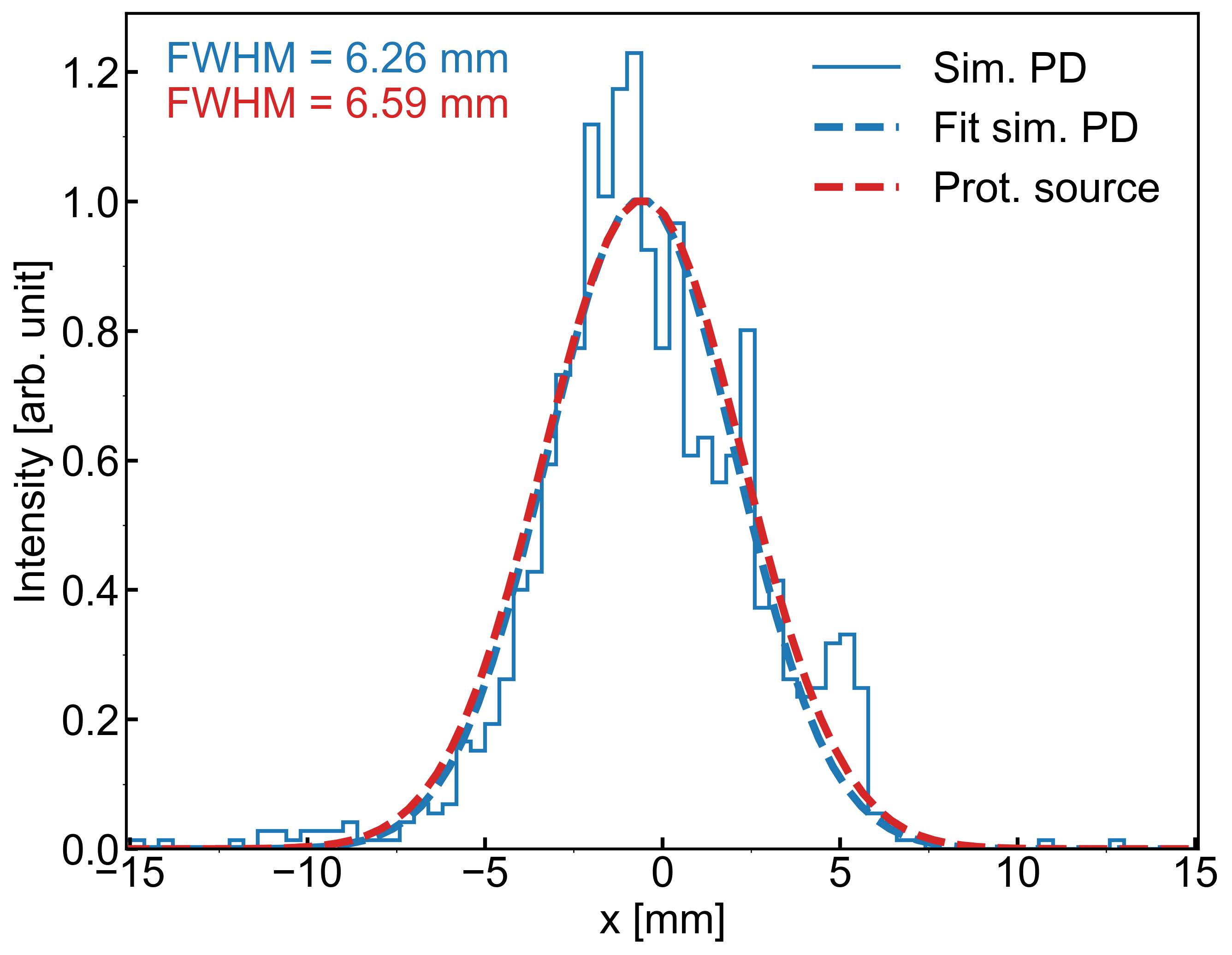}
\caption{Simulated  profiles computed at photodetectors position, considering their geometry (blue lines), compared with the horizontal profile of the proton source (red dashed line).}
\label{Profile_no_layer}
\end{center}
\end{figure}

\begin{figure}[htbp]
\begin{center}
\includegraphics[width=\linewidth]{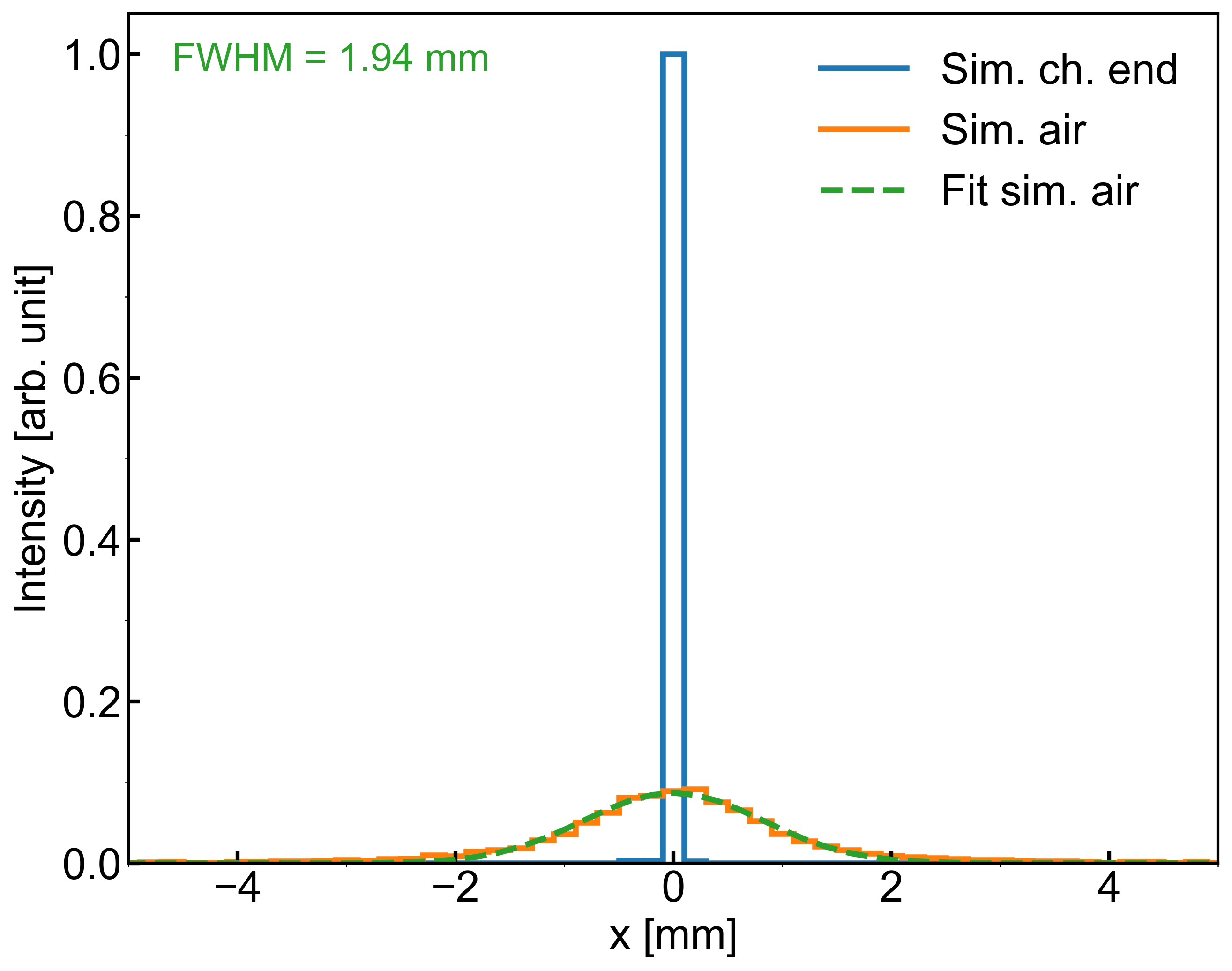}
\caption{Simulated profiles at the end of the channels (blue line) and after 1 mm of air (orange line) obtained using a point-like source (FWHM $=$ 141 \si{\micro m}) impinging on the central channel.  }
\label{air effect single channel}
\end{center}
\end{figure}

The device has been modeled using \textsc{gate} (v 9.1)\cite{jan2004gate,jan2011gate,sarrut2014review,sarrut2021advanced}, an open-access Monte-Carlo simulation software dedicated to medical imaging and radiotherapy, based on the \textsc{geant\scht{4}} toolkit \cite{agostinelli2003geant4}.
The backbone code used for our simulation has been retrieved from LUT Davis Model \cite{trigila2021optimization,LUTDavisGithub}, modified to include proton-matter interaction, geometry, and material discussed above. The simulation accounts for the photon generation from a proton beam, the photon transportation in the channels, and the final interface with air. The output of each simulation consists of two data frames: `Hits' and `Phase space'. The data frame `Hits' contains the position and the direction of each scintillated photon. The `Phase space', is a user-defined surface where the position and the direction of the outgoing particles (both protons and photons) are saved. To make the comparison with experimental results possible, the simulated proton beam has a Gaussian distribution with the same parameters as the one used during the experimental tests. The energy of the simulated protons is 300 mm Water Equivalent Thickness (WET), while the full-width half maximum (FWHM) is 7.72 mm. The direction of the beam has been set so that the beam impinges on the center of the active area. \\
Fig.~\ref{Profile_no_layer} shows the comparison between the profiles obtained at the PDs  and the proton source. The simulated profile is in good accordance with the expected one, showing an error of $5\%$ in the Full Width Half Maximum (FWHM); the non-gaussian deformations in the profile are simply due to the reduced statistics in the number of simulated particles (N = 100).\\ 
To quantify the effect of the air gap between the end of the polymeric waveguides and the PD array, which is inevitable in the first assembly presented in this paper, we simulated a point-like source having an FWHM of 141 \textmu m impinging on the central channel only. The results are reported in Fig.~\ref{air effect single channel}. While in the absence of an air gap, all the detected photons are those coming from the central channel, the introduction of the interface with air broadens the distribution of the photons (FWHM $\sim$ 1.90 mm) and reduces the number of collected photons by the PD in correspondence of the central channel by a factor of 10. This unwanted effect is obtained by simulating a flat surface of the channel extremity, which represents in first approximation the real case.\\

We can define the detection efficiency $\eta$ by taking into account different contributions: 
\begin{equation}
    \eta = \eta_{\textit{Scint}} \cdot \eta_{\textit{Waveguide}} \cdot \eta_{\textit{Air}} \cdot \eta_{\textit{PD}} 
\end{equation}
\noindent
where $\eta_{\textit{Scint}} = \frac{n_{\textit{gen}}}{n_{\textit{exp}}}$ is the ratio between the number of photons generated by the scintillation process and the one expected, computed as $n_{\textit{exp}} = Y \cdot E_{dep}$. \textit{Y} is the scintillation yield of the selected resin (9000 photons/MeV), and $E_{dep}$ is the energy deposited by protons and according to  the Bethe-Bloch formula considering the thickness of the channels (200 \textmu m) \cite{Kolanoski2020}. $E_{dep}$ depends on the impinging particle energy and intensity, $\eta_{\textit{Waveguide}}$ takes into account the transport efficiency of the photons inside the channels, \textit{i.e.}, it is the ratio between the photons arriving at the end of the channels and the total number of photons generated by scintillation. $\eta_{\textit{Air}}$ considers the effects of 1 mm of air between the channel end and the PDs and it is given by the number of photons collected by the PDs area over the ones at the end of the channels. Finally, $\eta_{\textit{PD}}$ is the quantum efficiency of the PD, defined as  $\eta_{\textit{PD}}=\frac{E\cdot R}{e}$, where \textit{E} is the photon energy, \textit{R} the PD responsivity, \textit{i.e.}, the ratio between the photocurrent and the incident optical power, and \textit{e} is the charge of the electron. In the examined case of a proton beam with an energy of 300 mm WET, $\eta_{\textit{Scint}}$ =  0.49, $\eta_{\textit{Waveguide}}$ = 0.16,  $\eta_{\textit{Air}}$ = 0.20 and $\eta_{\textit{PD}}$ = 0.59 for photon energy of 423 nm, which is the peak emission of the scintillating resin used to fabricate the detector's active area. Thus, the total efficiency amounts to 1.2 $\%$. $\eta_{\textit{Air}}$ changes with the PDs geometry and their distance with respect to the channel end.

\subsection{\label{sec:physical} Microfabrication Process}

\begin{figure}[htbp]
\begin{center}
\includegraphics[width=\linewidth]{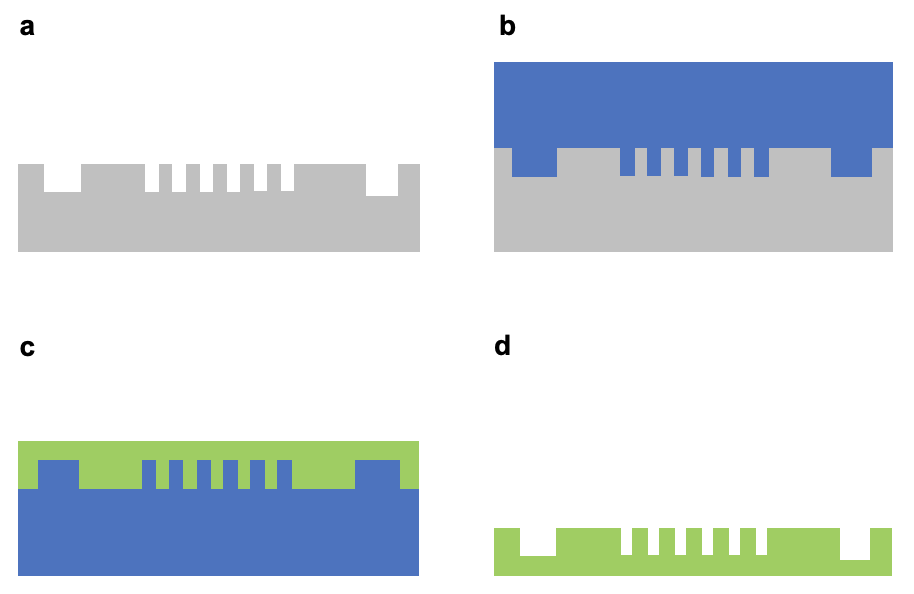}
\caption{Main steps of the fabrication process: a) channels etching into a silicon wafer, b) PDMS mold made using the silicon master, c) pouring of the scintillating resin into the PDMS mold, d) the final active area made in resin is a replica of the silicon master.}
\label{Fig: PF}
\end{center}
\end{figure}

\begin{figure}[htbp]
\begin{center}
\includegraphics[width=\linewidth]{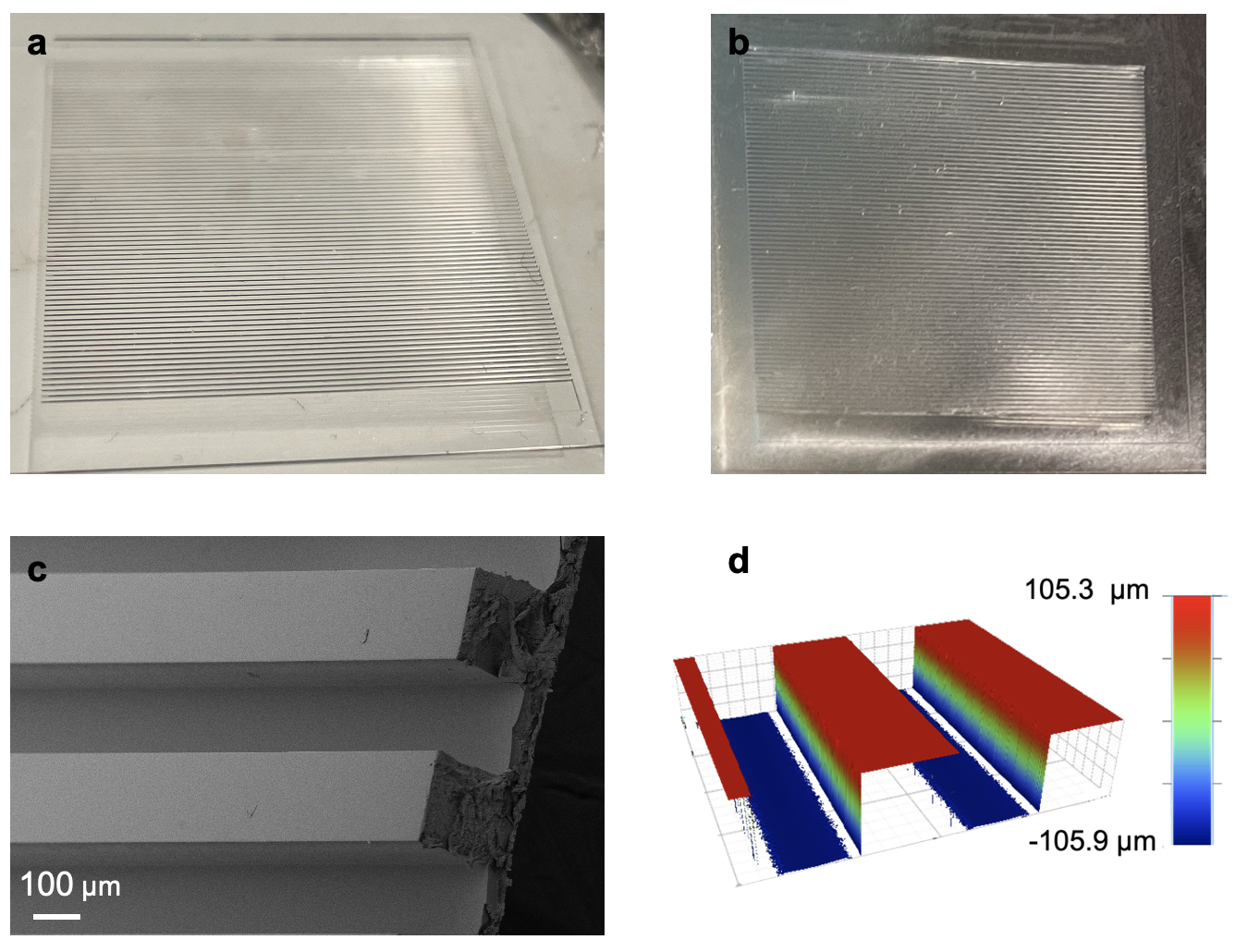}
\caption{a) PDMS mold realized using a silicon master patterned with 200 \textmu m-width channels. b) Microchannels made by scintillating resin, obtained by demolding the resin from PDMS molds, after a PFTOS surface treatment. c) and d) Scanning electron microscope (SEM) image and 3D measurement of the resin active area geometry taken by an optical profilometer, showing smooth vertical walls and a very accurate reproduction of the PDMS mold.}
\label{Fig: profilometer}
\end{center}
\end{figure}

The realized beam profiler consists of microchannels made by scintillating resin (EJ-290 by Eljen Technology) coupled to a readout system, in our case a photodetector array (S8865-128 by Hamamatsu). The micrometric dimensions of the channels together with the photodetectors acquisition properties enable high spatial resolution and high frame-rate readings. In such systems, once the particle beam (\textit{e.g.}, protons, ions, etc.) hits the scintillating materials, light is generated due to the scintillation process. The photons are then partially guided to the photodetector system thanks to the Total Internal Reflection (TIR) phenomenon and collected by the photodetector array. The latter converts the light into a current, which is proportional to the number of photons detected by each photodetector. In such a way, it is possible to reconstruct the one-dimensional beam transverse profile with minimal perturbation of the beam itself. A schematic of the setup is depicted in Fig.~\ref{sketch}.\\
The scintillating microchannels have been realized starting from silicon masters patterned with microchannels having a width of 200 \textmu m and a pitch of 400 \textmu m. We fabricated silicon masters by standard microfabrication techniques, \textit{i.e.}, optical lithography for patterning the channels, followed by deep reactive ion etching (DRIE) (see Fig.~\ref{Fig: PF}a). After the DRIE, silicon walls feature a typical undulation, known as the `scalloping effect`, which is due to a slight in-plane etching of silicon. To remove the scalloping effect, a layer of 2 \textmu m of silicon oxide was grown on the etched channels and etched afterward by buffered hydrofluoric acid (BHF) \cite{Maoddi2014}. Having smooth walls is crucial for avoiding significant photon losses during their transmission toward the photodetectors.\\
The silicon masters were used to make Polydimethylsiloxane (PDMS) molds (see Fig.~\ref{Fig: PF}b). The pattern on the silicon master was replicated with high accuracy (roughness better than 500 nm) on PDMS molds (see Fig.~\ref{Fig: profilometer}a), which were filled with the scintillating resin thereafter (see Fig.~\ref{Fig: PF}c). The final active area made in resin is a replica of the original silicon master (see Fig.~\ref{Fig: PF}d). The resin is composed of three parts. Part A contains a partially polymerized plastic scintillator, \textit{i.e.}, oligomers of vinyl toluene (VT). Part B is composed of VT monomers, 2,5- Diphenyloxazole (PPO), 1,4-Bis(5-phenyl-2-oxyzole) benzene (POPOP), and 2,6-Dit-tert- butyl-p-cresol (DBPC). PPO and POPOP are the primary and secondary fluorophores operating as wave shifters. The wave shifters are added to convert the nonradiative ionization radiation produced by polyvinyl toluene to lower energy photons (blue or green) that are detectable with the photodetector array. Part C contains Lauroyl peroxide, a thermal initiator, added for polymerization and crosslinking of vinyl toluene.\\
The resin is viscous before polymerization which takes approximately 3 hours at 80 °C. Once it is polymerized, it can be easily demolded from the polymeric mold: the result is a self-standing scintillating area. To facilitate the demolding process, a surface treatment was performed on the polymeric molds. It consists of exposing the mold to the oxygen plasma, followed by a silane (Perfluorooctyltriethoxysilane, PFTOS) coating deposited in gas phase. Fig.~\ref{Fig: profilometer}b shows the resin active area after the demolding; it is worth noting that a residual thin layer of resin connects the channels. The characterization of the final active area has been performed with a Scanning Electron Microscope (SEM) (see Fig.~\ref{Fig: profilometer}c) and an optical profilometer (see Fig.~\ref{Fig: profilometer}d), which confirms the depth of the channels to be $\sim$ 211 $\pm$ 1 \textmu m. The channel width and pitch are dictated by the silicon master which is obtained by standard photolithography and etching processes. The thickness of the residual thin layer was measured to be $\sim$ 80 \textmu m.  A more detailed description of the fabrication process is reported in the Supplementary Information.\\
The main advantage of this fabrication method is that it is possible to obtain geometrically accurate and low-cost detectors, which can be easily replaced if a degradation due to radiation damage occurs. In addition, the achievable resolution is higher than the standard scintillating fiber-based detectors currently used for beam diagnostics and experiments. Moreover, the aforementioned detectors could cover a wide range of applications because they can be designed to suit in principle all types of protons or heavy ion medical accelerators, namely cyclotrons, synchrotrons, and linacs, but they could also be used for dosimetry or X-ray imaging. 

\subsection{\label{sec:electronics} Readout and Control}

\subsubsection{Photodiode Array}
The scintillating light exiting each detector channel needs to be carefully converted into an electronic signal to obtain an accurate profile. For this purpose, a linear array of 128 photodiodes with integrated amplifiers have been chosen (S13885-128, Hamamatsu). In particular, the PD pitch is 400 \textmu m, each PD is 300 \textmu m large and its height is 600\textmu m. Moreover, the PDs work in an extended visible spectrum range from 200 nm to 1000 nm, with a peak sensitivity at $\lambda$ = 720 nm. At the peak wavelength of the scintillation photons (423 nm) the approximate PD sensitivity is 0.26 A/W.\\
One of the biggest advantages of using commercial PD arrays is the presence on board of an integrated charge amplifier array followed by a clamp and hold circuit, allowing a serial output and a relatively simple digital interface, consisting of only a clock (CLK) and a line aimed to set the desired integration time of the charge integrators connected to each PD. 
The achievable profile rate, which depends on the CLK frequency $f_{CLK}$ and on the number of PD $N_{PD}$, is expressed in the Eq.~\ref{eq:tprofile}.
\begin{equation}
\label{eq:tprofile}
f_{Profile} \leq f_{CLK}/(16.5+4\cdot N_{PD} )
\end{equation}
Considering that the maximum allowed clock frequency is 4 MHz, the maximum profile rate can be as high as 7568 Hz, which is considerably higher than commonly used scientific cameras and largely fulfills the typical needs of a beam profiler in a DC or pulsed particle accelerator;
The minimum integration time can be as low as 4.5 \textmu s if the maximum CLK frequency of 4 MHz is used.
The maximum integration time is limited by the saturation of the PD analog output. The dark state output voltage is typically 2.5 V (normally high) and the PDs, when illuminated, generate a train of inverted pulses whose amplitude is proportional to the integrated light. The voltage of each PD output can go as low as 0.7 V when it reaches saturation. 
The photoresponse non-uniformity declared by the manufacturer can be as high as $\pm$10\%, which would directly translate into a profile deformation. For this reason, an online calibration tool using UV light has been developed.\\
The PD array is mounted on a G10 glass-epoxy printed circuit board (PCB). The electronics and the wire bonds are protected with resin. The PCB  presents four 2.2 mm diameter mounting holes precisely machined with respect to the position of the PD array. With tolerances in the order of $\pm$0.2 mm, an accurate alignment procedure must be in place during the assembly of the PD array with the detector, in order to maximize the photon signal coming from each scintillating channel.

\subsubsection{Controller}
The second electronic block of the system is a controller that has two main functions: i) to generate all needed supply voltages for the PD array and ii) to provide the analog and digital interface to it to obtain the digitized profile with the desired settings. The core of the custom-developed board is a programmable microcontroller module (Nucleo-G431KB, STMicroelectronics), which is powered by a standard micro-USB cable with 5 V. The Nucleo board generates a 3.3 V (max 500 mA) which is used to power the PD array. For what concerns the 2.5 V, needed for the analog circuitry of the PD array integrated circuit, a low noise low dropout regulator is used (LP5907, Texas Instruments). The analog signal $V_{Out}$ from the PD array is buffered and takes two parallel paths: the MCU analog input and a coaxial SMA connector for acquisition into another system. The MCU in fact includes an ADC of 12 bits capable of reaching a sample rate of 4 MHz and presenting also an external trigger capability, using the trigger signal coming from the PD array. The clock and the integration timeline are programmable with the MCU firmware, allowing maximum flexibility according to the use case of the measurements. The controller PCB presents also an external trigger connector, such as to measure and acquire a profile synchronous with an experiment event. Finally, the serial interface peripheral of the MCU is connected to the PCB to program a data transfer across a field cable if necessary.

\subsubsection{UV Led calibration tool}
\begin{figure}[htbp]
\begin{center}
\includegraphics[width=\linewidth]{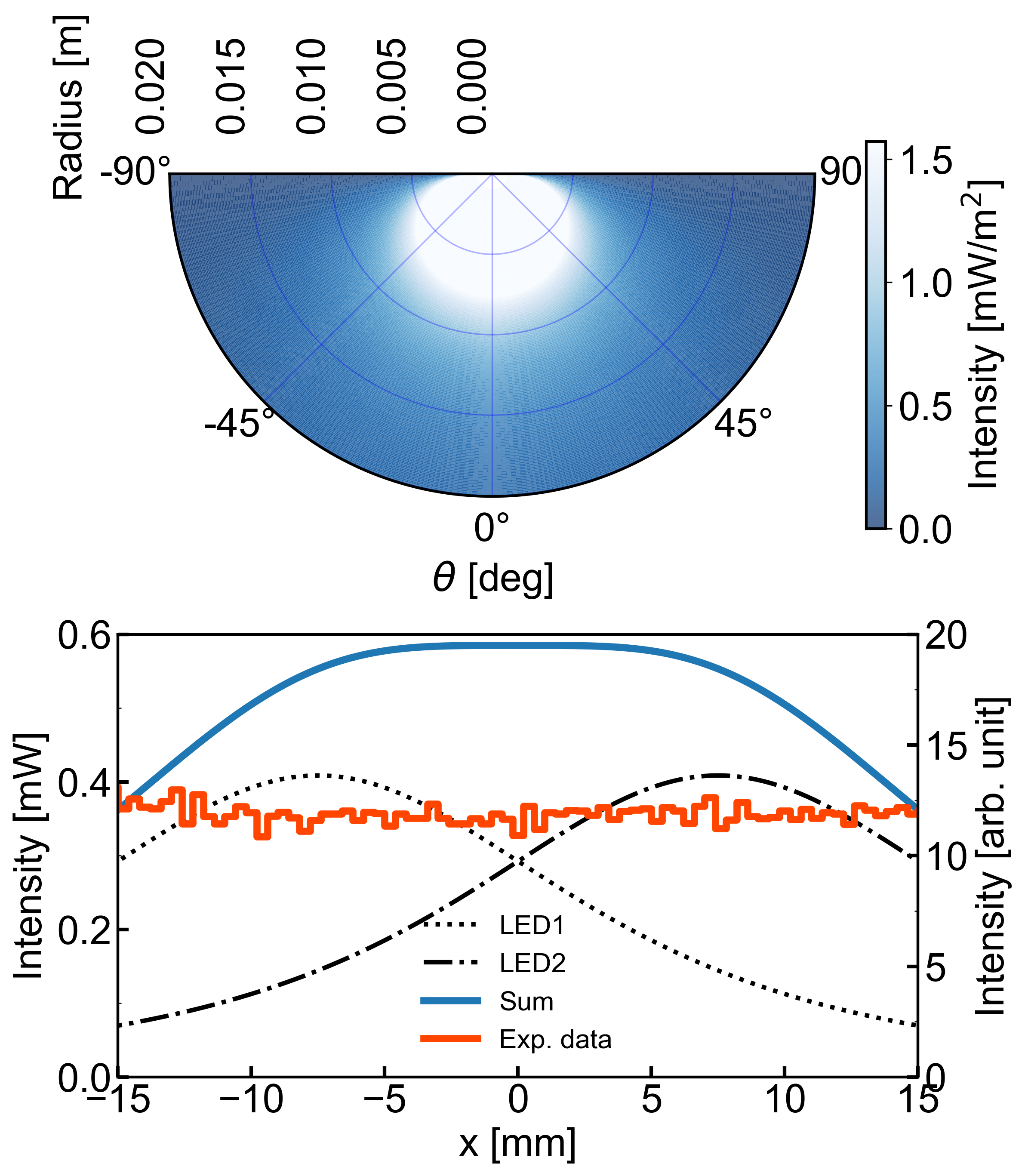}
\caption{Lambertian emission of the UV Led using an emitted power of 2.6mW at $I_f$ = 20mA (Top) and theoretical luminous intensity at the detector surface due to both LEDs and experimental profile (Bottom). }
\label{UV}
\end{center}
\end{figure}

An additional PCB has been developed with the main purpose of illuminating the detector's active area with UV-C light. Despite this kind of source is not able to generate true scintillation in the material, it is still capable of being absorbed by the fluors of the scintillator, which would re-emit the light in the visible range. This method turned out to be a convenient tool to determine the overall detector yield, which is a combination of channel fabrication, detector-PD array misalignment, and PD array non-uniformity. Moreover, as suggested in \cite{Wetzel_2022}, UV light might help in the radiation damage recovery process of the scintillator. Two LEDs spaced by 15 mm (VLMU35CB2-275-120, Vishay) emit narrow UV-C radiation with a peak wavelength at 273 nm; aligned with the periphery of the detector, the emitted UV light reaches the active area thanks to the angle of the half intensity of $\pm60^{\circ}$. Fig.~\ref{UV} shows the Lambertian emission of one single LED in a polar plot and the theoretical luminous intensity resulting at the detector surface in correspondence of the LEDs, in the case when both LEDs are lit. In orange is reported the experimental profile, which is mostly flat across the entire detector width (30 mm). The presence of two LEDs aimed at the potential illumination of a large area detector with dimensions up to 80 mm.

\section{\label{sec:experimental} Experimental validation}
\begin{figure}[htbp]
\begin{center}
\includegraphics[width=3.2in]{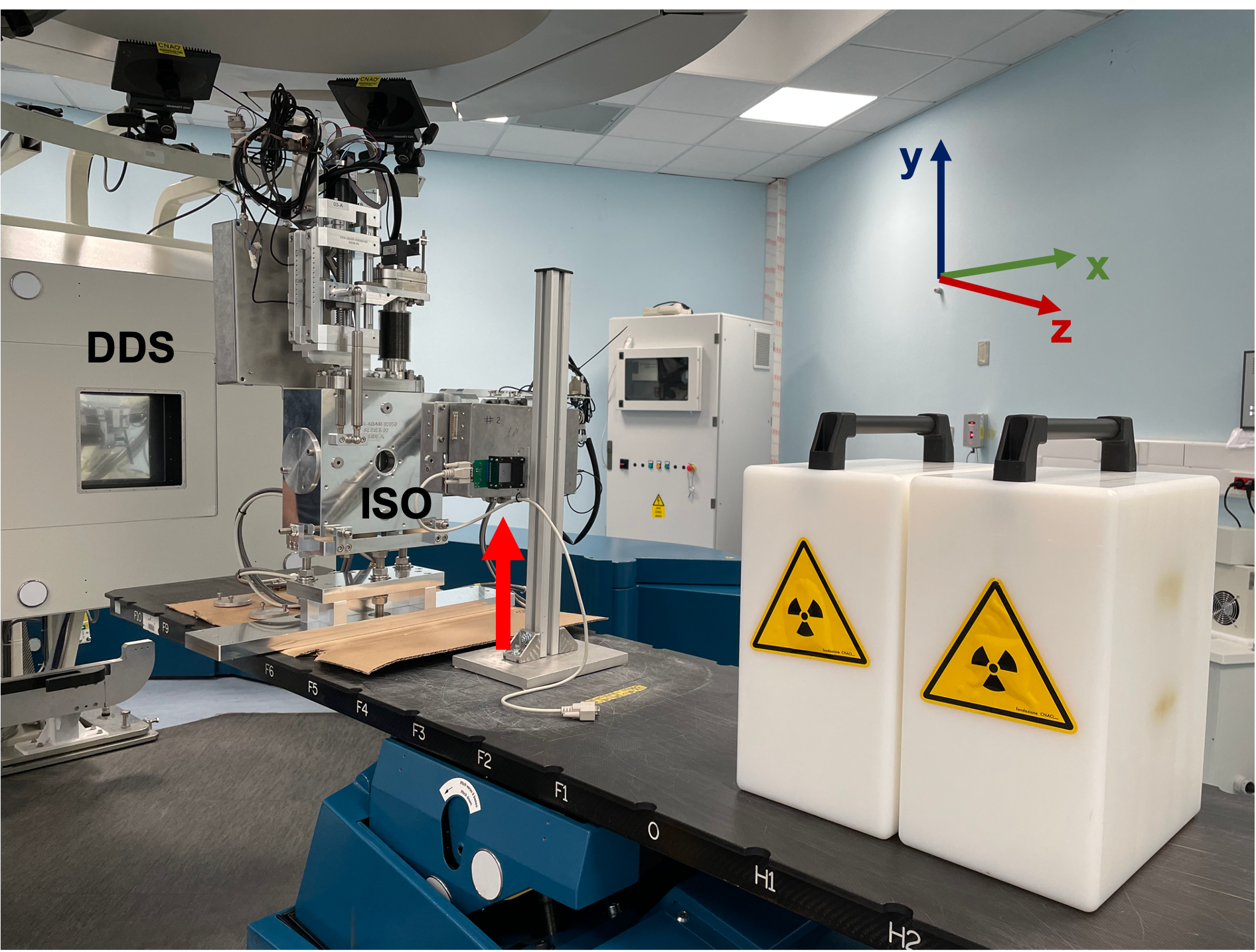}
\caption{Treatment room at CNAO. The red arrow indicates the device, which is placed on the patient's bed $\sim$ 55 cm from the isocenter. The white boxes at the end of the bed are full of water and are meant to stop the beam. The two reference detectors (DDS and ISO) are also marked.}
\label{Beamline}
\end{center}
\end{figure}

The scintillator has been tested on a proton beam at the National Centre for Oncological Hadrontherapy (CNAO) in Pavia, Italy \cite{ROSSI2015333}. Each of the three available treatment rooms at CNAO can deliver to patients mainly protons or carbon ions with an energy ranging from 62.73 MeV (30 mm WET) to 228.57 MeV (320 mm WET) and 115.23 MeV/u (30 mm WET) to 398.84 MeV/u (270 mm WET) respectively  \cite{Mirandola}. The beam FWHM at the isocenter varies as a function of energy from 22 mm to 7 mm in both transverse planes. In the performed experiments the protons were sent to the isocenter on their nominal trajectory and, after traversing the device under test, they were dumped in a water tank. In each extraction, lasting typically 1 s, up to $3\cdot 10^9$ protons can be directed to the patient. Each treatment room is equipped with a Dose Delivery System (DDS), capable of monitoring and controlling the beam delivery according to the treatment plan. In particular, at CNAO, the required beam charge is constantly monitored and the beam being extracted from the synchrotron is interrupted when the desired charge is delivered in the treatment room at a given energy. During the experiments, the charge was set to $5\cdot 10^8$ protons. The DDS is composed of a strip ionization chamber with a strip pitch of 1.65 mm, plus an integral plane ionization chamber to monitor the charge. In the results presented in this section, an additional set of reference profiles at the isocenter (ISO) has been also considered. Those data have been taken prior to the experiment with a dedicated ionization chamber with a strip pitch of 1 mm. The DDS is permanently installed 87.35 cm upstream with respect to the isocenter, in the air, just after a vacuum window. Our prototype has been installed on the patient bed approximately 55 cm downstream of the isocenter, on a support stand (see Fig.~\ref{Beamline}). Due to the limited time available and the mechanical setup, only the vertical transverse plane was measured. The profile measurements were triggered using the timing signal of extraction start and the PD signals were integrated for 10 ms. The profiles were acquired on an oscilloscope, which was remotised in the medical treatment room and the profiles were post-processed in Python.

\begin{figure*}[htbp]
\begin{center}
\includegraphics[width={\textwidth}]{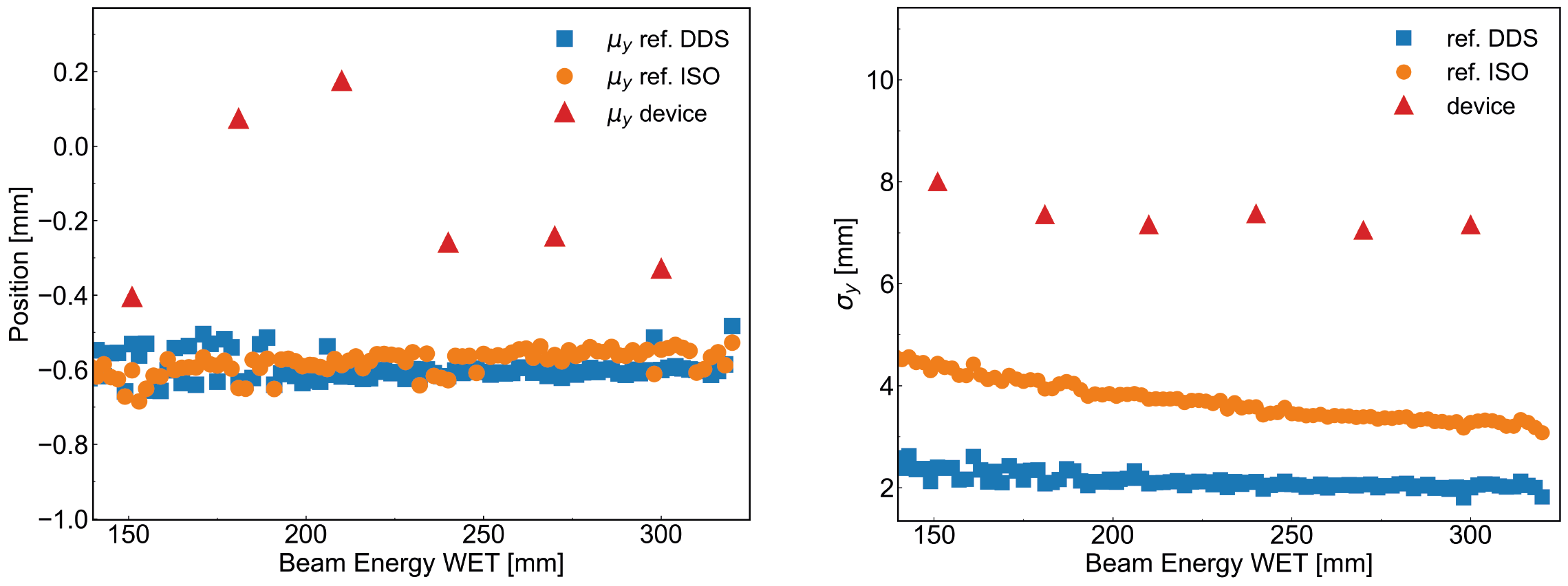}
\caption{Summary of the experimental results on CNAO proton beams for different energies.}
\label{CNAO_musigma}
\end{center}
\end{figure*}

The detector successfully measured its first proton profiles, demonstrating a high sensitivity. In fact, for proton energies lower than 141 mm WET the measured profiles saturated. The profiles were visible until the highest tested energy of 300 mm WET. Since the typical extraction lasts 1 s and the charge was fixed at $5\cdot 10^8$ protons, we can deduce that our beam profiler was measuring profiles of approximately 5 million protons.\\
The beam spot center of mass (COM) and standard deviation as a function of energy are shown in Fig.~\ref{CNAO_musigma}; the reference profiles at the DDS and the ISO have been previously measured and stored by CNAO experts. For what concerns data processing, since the profiles might present tails or some pedestal that can heavily affect the computation of the center of mass and transverse size, the raw profile amplitudes have been zeroed for values below $5\%$ of the maximum. It can be noted that, while the COM is relatively constant at various energies, the beam size increases from DDS to the ISO due to the Multiple Coulomb Scattering (MCS) through the vacuum window, the DDS material, and the path in the air. Moreover, as expected, the beam spot opening is more pronounced for low energies. The COM measured with our device differs 700 \textmu m from the reference ones, which can be explained by the manual alignment, which was made only using two orthogonal lasers indicating the x-y and y-z planes (see axes in Fig.~\ref{Beamline}). For what concerns the beam transverse size, the detector measured systematically a beam much larger than expected. Computing the contribution of the MCS using Highland \cite{LYNCH19916} approximation, the beam transverse sigma should increase only between 70 and 30 \textmu m from low to high energy with respect to that shown in Fig.~\ref{CNAO_musigma} at ISO. Thus the beam profile is indeed broadened by additional fabrication and geometrical factors, as discussed in the next section.

\section{Discussion}
\begin{figure}[htbp]
\begin{center}
\includegraphics[width=\linewidth]{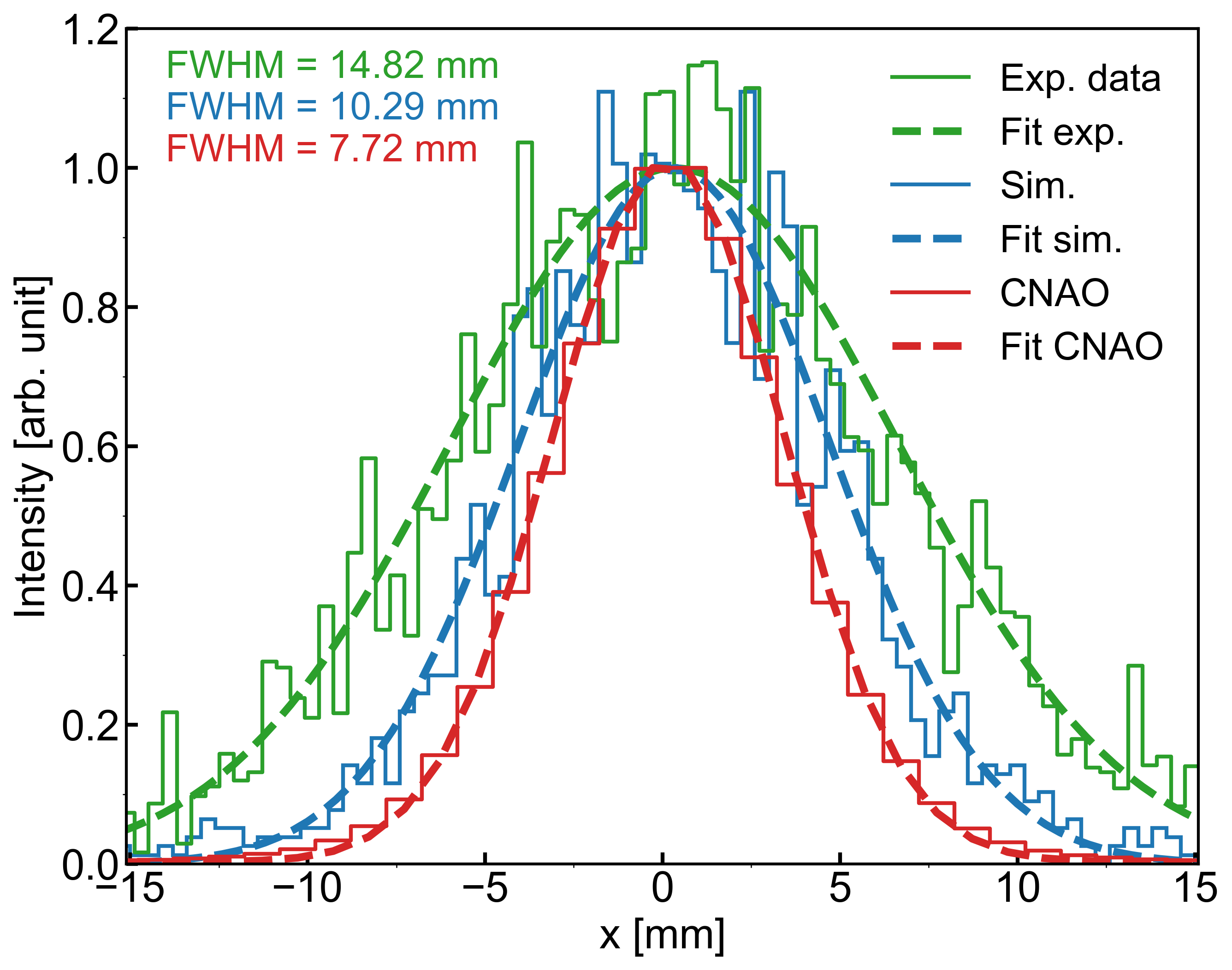}
\caption{Experimental (green) proton beam profile at 300 mm WET energy, compared with the reference profile at ISO (red) and simulated (blue).}
\label{Profile_si_layer}
\end{center}
\end{figure}

 We implemented the geometry of the real detector in the simulation model to quantify the contribution of the residual thin layer of scintillating material that binds together the microfabricated waveguides and the 1 mm air gap between the waveguides end and the PDs, in order to study the profile broadening. Fig.~\ref{Profile_si_layer} shows the reference beam profile (CNAO) together with the measured and simulated one. The latter reveals an FWHM $\sim$ 35$\%$ larger than the nominal one. However, the experimental profile results to be $\sim$ 44 $\%$ broader than the simulated one, which could be due to additional factors such as the roughness and defects at the channel end's surface, the vertical alignment between the waveguides and the photodiodes and a too large air gap. 
In terms of overall detector efficiency, using the beam properties and the detector geometry of Fig.~\ref{Profile_si_layer}, we computed and simulated the different efficiency contributions introduced in Section \ref{sec:modelling}. The $\eta_{\textit{Scint}}$, due to the presence of the residual thin layer, increases to the value of 0.60 (0.49 with no layer), while $\eta_{\textit{Waveguide}}$ didn't change. The resulting total efficiency amounts to 1.5$\%$. 

\begin{figure}[htbp]
\centering
\includegraphics[width=\linewidth]{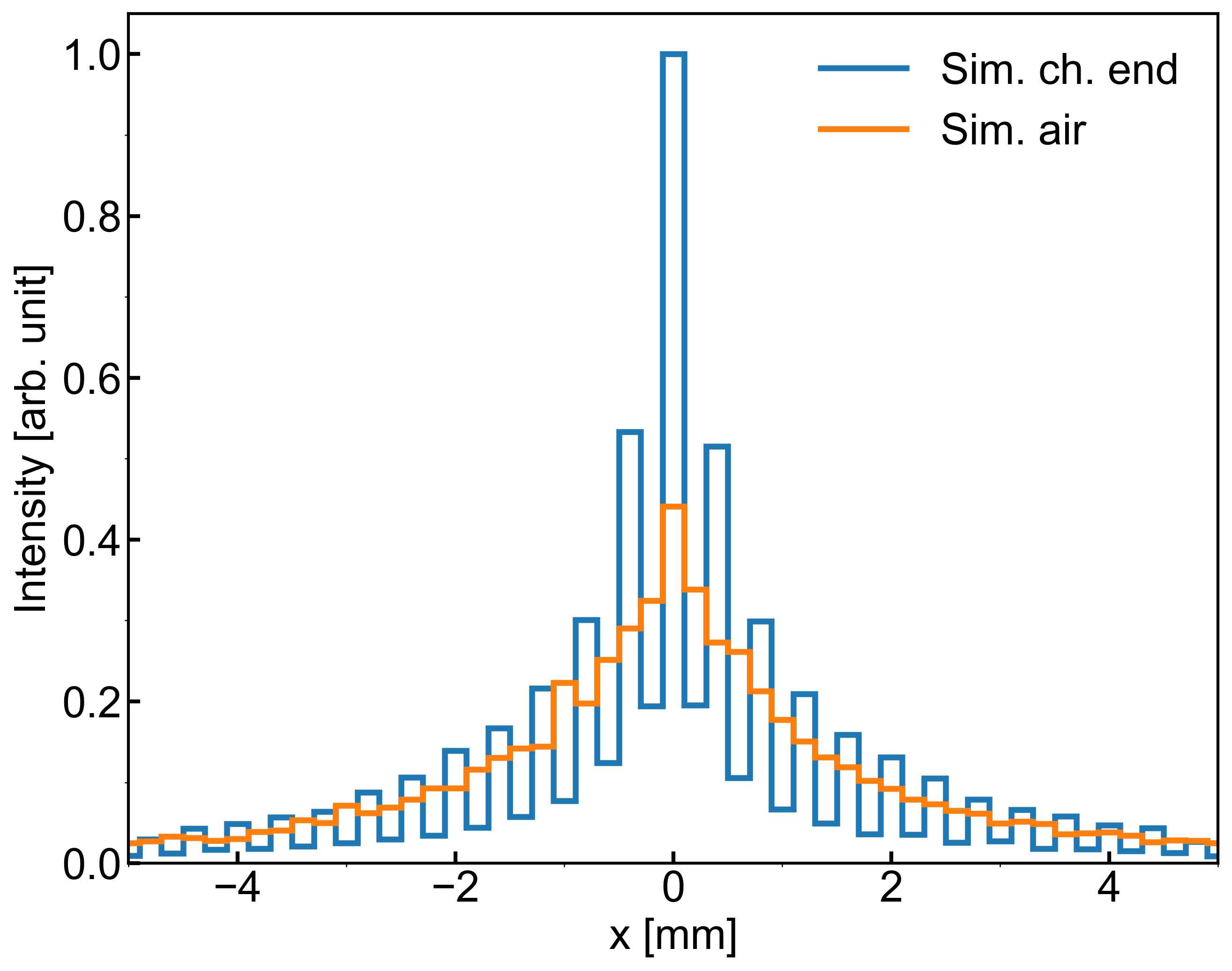}
\caption{Simulated beam profiles of detector plus the thin residual layer linking all waveguides with readout at channels end (blue line) and after 1 mm of air (orange line) with a point-like source with an FWHM of 141 \si{\micro m}.}
\label{air_end_a}
\end{figure}

\begin{figure}[htbp]
\centering
\includegraphics[width=\linewidth]{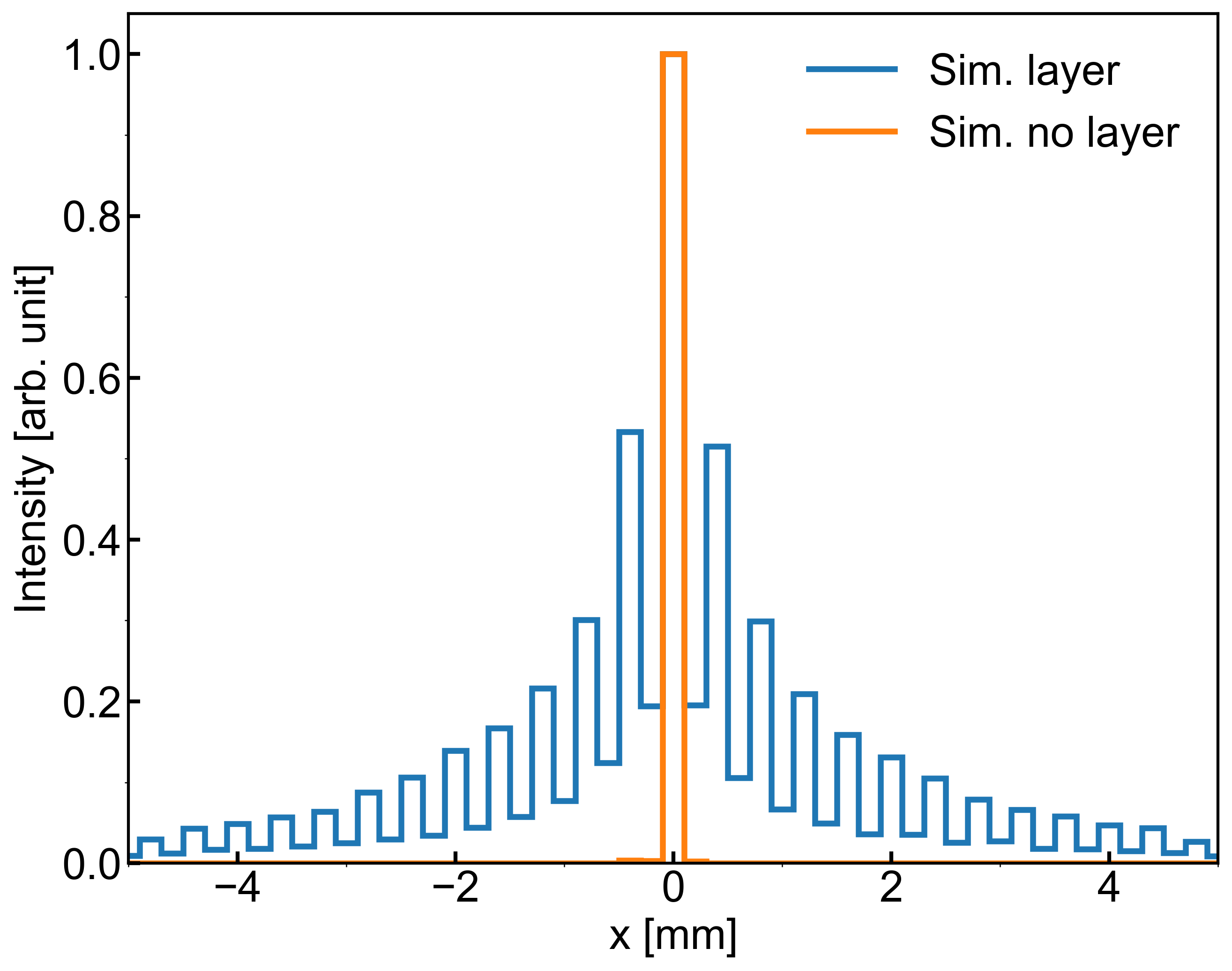}
\caption{Comparison between simulated profiles at the end of the channels of a device with (blue curve) and without the thin layer linking all the microfabricated waveguides (orange curve) with a point-like source with an FWHM of 141 \si{\micro m}.}
\label{air_end_b}
\end{figure}

In order to further investigate the effect shown in Fig.~\ref{Profile_si_layer}, we evaluated the effect of 1 mm of air gap between the channel end and the PDs simulating a point-like source (FWHM = 141 \textmu m) impinging on the central channel of the device, considering this time also the residual thin layer. As expected, the combined effect of the layer and the air gap spread, even more, the photons from scintillation with a non-gaussian distribution covering $\pm$ 5 mm around the stimulated channel (see Fig.~\ref{air_end_a}).  
The layer that links the waveguides is the main factor contributing to the photons' cross-talk between adjacent channels, resulting in the profile broadening. In fact, without the layer, the cross-talk is only due to the air gap, which spreads the photons on a region of $\pm$ 2 mm around the stimulated channel (see Fig.~\ref{air_end_b}). However, this layer serves as a mechanical support for the channels and maintains the whole structure together. 

\section{Conclusion}
In this work, we presented the development of a novel beam profiler obtained by casting scintillating resin on a microfabricated structure. The proof of concept detector covers an area of 30 x 30 mm$^2$ with an array of square section scintillating waveguides of 200 \textmu m spaced by 400 \textmu m. It is mechanically self-sustaining and it exploits the surrounding air/vacuum refractive index to guide the scintillating photons along its channels to a photodiode array, where one-shot profiles can be measured at a rate as high as 7 kHz. A prototype of the detector, holder, and readout electronics was used first under UV illumination and measured successfully its first profiles on a real proton beam used for therapy at CNAO. The preliminary measurements presented in this work showed great sensitivity across all energy ranges (70-230 MeV) using only a few million protons to get a profile. However, the thin layer of resin linking all channels and the air gap between the channel ends and the PD are the main contributors to the profile broadening. They should be addressed in the next fabrication iterations. Other aspects must be improved in the future, \textit{e.g.}, the coupling between the active area and the PD array to minimize the photon losses, the demolding process, and the surface quality of the channel extremities. The detector is currently being built with double resolution (pitch 200 \textmu m) and improved mechanical properties.

\section*{Acknowledgements}

The authors are deeply grateful to CNAO for the availability and the help in setting up the preliminary measurements with the beam; a particular thanks to Claudio Viviani for the assistance during the experiment. Moreover, we would like to thank Theodore Rutter (ADAM SA) for the mechanical aspects, Maurice Haguenauer (CERN) for the fruitful discussions, and all personnel of the Center of MicroNanoTechnology (CMi) of EPFL for the technical support during the detector fabrication steps.

\newpage
\bibliography{apssamp}
\appendix
\section{Supplementary Information}

\begin{figure*}[htbp]
\begin{center}
\includegraphics[width=5.8in]{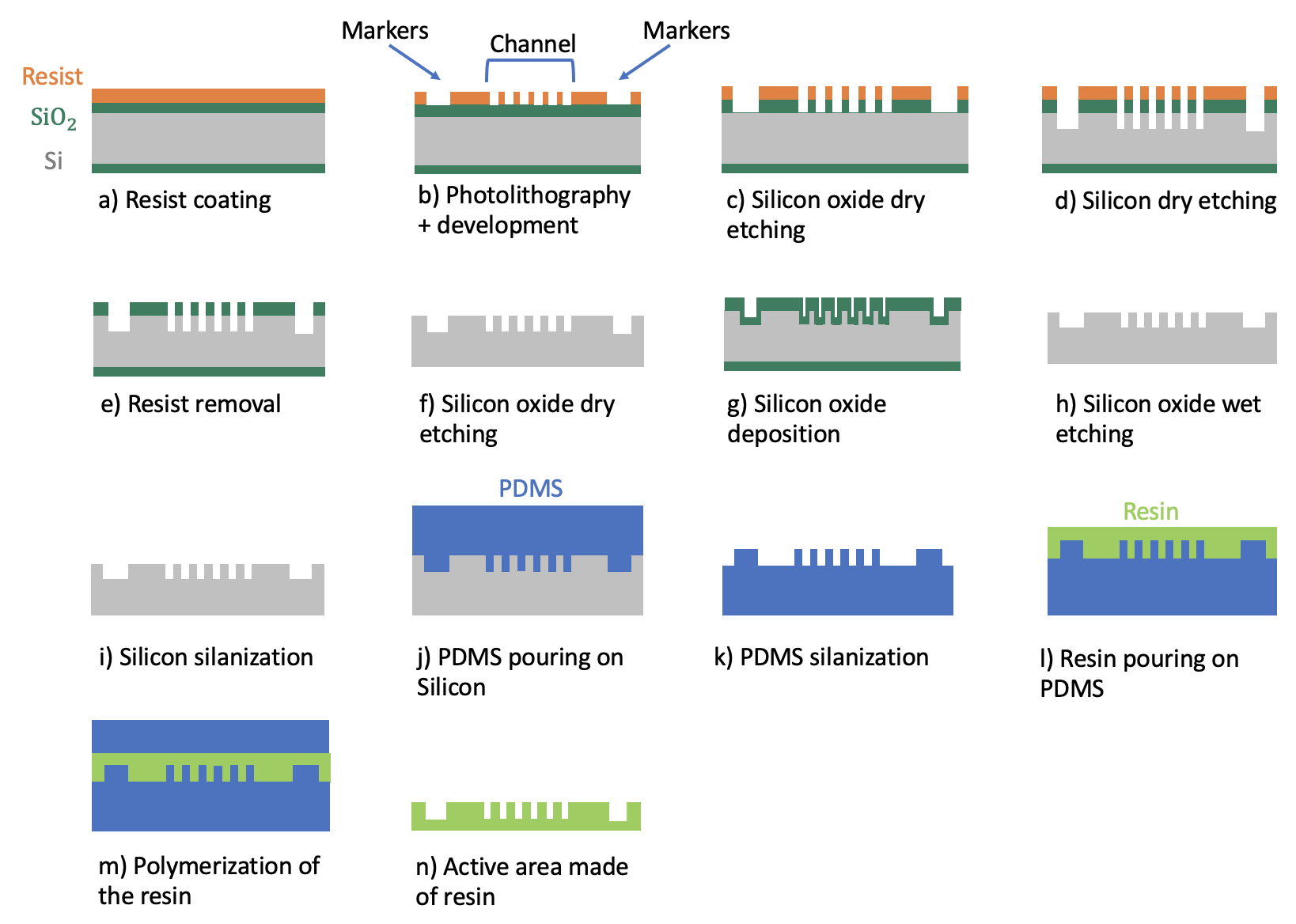}
\caption{Process flow of the self-standing active area made by scintillating resin.}
\label{Fig: PF_resin}
\end{center}
\end{figure*}

\maketitle
The fabrication process of the resin-based self-standing active area consists of the preparation of the silicon master, which is used to fabricate the PDMS mold. The latter, in turn, is utilized for the realization of the active area made of resin.\\
The process starts with silicon master manufacturing, which involves standard microfabrication steps. The first step is the coating of a silicon wafer with a photoresist layer (5 \textmu m-thick AZ10XT-02) (see Fig. \ref{Fig: PF_resin}a). The wafer is double-side polished, 380 \textmu m-thick, covered with 2 \textmu m of silicon oxide (SiO$_{\mathrm{2}}$). The following steps consist of photolithography to pattern the design onto the wafer (see Fig. \ref{Fig: PF_resin}b). In this case, the design comprises a series of channels held together by a frame. The fabricated active areas have a channel width of 200 \textmu m with a pitch of 400 \textmu m to match the pitch of the PD array. The channels are then etched by 200 \textmu m. First the SiO$_{\mathrm{2}}$ is etched with a mixture of 4F$_{\mathrm{8}}$/H$_{\mathrm{2}}$/He using an Inductively Coupled Plasma Reactive Ion Etching (ICP-RIE) see Fig. \ref{Fig: PF_resin}c). The silicon etching is then carried out with a Deep Reactive Ion Etching (DRIE) tool (see Fig. \ref{Fig: PF_resin}d).
The DRIE method is commonly utilized to create structures with high aspect ratios and vertical sidewalls. This is accomplished by utilizing two different fluorine-based plasmas, SF$_{\mathrm{6}}$ and C$_{\mathrm{4}}$F$_{\mathrm{8}}$. SF$_{\mathrm{6}}$ is responsible for chemically and physically attacking the silicon, resulting in vertical etching due to the applied electric field. On the other hand, C$_{\mathrm{4}}$F$_{\mathrm{8}}$ passivates the sidewalls to prevent horizontal etching. However, the so-called `scalloping effect', \textit{i.e.}, ripples ranging from 100 nm to 400 nm, may appear on the vertical walls after the silicon etching. This can impact photon transport due to the comparable wavelength. To remedy this, a layer of SiO$_{\mathrm{2}}$ is grown on top of the silicon wafer through wet oxidation. As the SiO$_{\mathrm{2}}$ layer grows, it smooths out the interface between Si and SiO$_{\mathrm{2}}$, resulting in a very smooth silicon surface. This guarantees optical reflections during photon transportation. \\
Thus, after the removal of the resist with the oxygen plasma (see Fig. \ref{Fig: PF_resin}e) the SiO$_{\mathrm{2}}$ layer is dry-etched see Fig. \ref{Fig: PF_resin}f), and a new layer of 2 \textmu m of SiO$_{\mathrm{2}}$ is grown by wet oxidation see Fig. \ref{Fig: PF_resin}g). Finally, the SiO$_{\mathrm{2}}$ layer is removed through BHF buffered hydrofluoric acid, which does not erode silicon (see Fig. \ref{Fig: PF_resin}h). 
Once the silicon master is ready, it can be used to make PDMS molds.  Chlorotrimethylsilane (TMCS) is evaporated on the silicon surface (see Fig. \ref{Fig: PF_resin}i) to facilitate the demolding process.
Preparing the PDMS involves mixing the PDMS base with the curing agent in a standard ratio of 10:1 or a ratio of 10:0.5 to make it more flexible during demolding. The mixture is then degassed in a desiccator to remove any air bubbles. The PDMS is then poured onto the silicon wafer (see Fig. \ref{Fig: PF_resin}j) and placed in the desiccator again. The silicon wafer with PDMS on top is then kept at 80°C for at least two hours. After curing, the PDMS is stable and can be carefully demolded from the silicon wafer. This process transfers the silicon features onto the PDMS with nanometric accuracy.\\
To prevent resin from entering the PDMS molds, the PDMS is subjected to oxygen plasma and silanized with perfluoro-terminated silane (PFOTS) (see Fig. \ref{Fig: PF_resin}k), which increases the hydrophobicity of the PDMS surface by creating a fluorinated monolayer. This prevents bonds from forming between the resin and the PDMS molds due to the presence of high electronegative atoms. Once the resin is poured onto the mold (see Fig. \ref{Fig: PF_resin}l), it is polymerized, which can take 14 days in a water bath at 47°C but can be achieved in just 4 hours by keeping the resin in an oven at 80°C. A piece of silanized PDMS (with no pattern on top) together with weight is added to the PDMS mold during the resin polymerization (see Fig. \ref{Fig: PF_resin}m) to minimize the resin layer between the channels. Despite this, a thin resin layer may still remain after the demolding (see Fig. \ref{Fig: PF_resin}n).

\end{document}